\documentclass[iicol,pdflatex,sn-mathphys-num]{sn-jnl}

\usepackage{graphicx}%
\usepackage{multirow}%
\usepackage{amsmath,amssymb,amsfonts}%
\usepackage{amsthm}%
\usepackage{mathrsfs}%
\usepackage[title]{appendix}%
\usepackage{xcolor}%
\usepackage{textcomp}%
\usepackage{manyfoot}%
\usepackage{booktabs}%
\usepackage{algorithm}%
\usepackage{algorithmicx}%
\usepackage{algpseudocode}%
\usepackage{listings}%

\theoremstyle{thmstyleone}%
\theoremstyle{thmstyletwo}%

\theoremstyle{thmstylethree}%

\begin{document}

\title[Article Title]{Thermodynamic Non-Uniformities Behind Incident and Reflected Shocks in a Single-Diaphragm Shock Tube}


\author[1]{\fnm{Touqeer Anwar} \sur{Kashif}}\email{touqeer.kashif@kaust.edu.sa}
\author*[1]{\fnm{Janardhanraj} \sur{Subburaj}}\email{janardhanraj.subburaj@kaust.edu.sa}
\author[1]{\fnm{Aamir} \sur{Farooq}}\email{aamir.farooq@kaust.edu.sa}




\affil[1]{\orgdiv{Mechanical Engineering, PSE Division}, \orgname{King Abdullah University of Science and Technology (KAUST)}, \city{Thuwal}, \postcode{23955-6900}, \country{Kingdom of Saudi Arabia}}




\abstract{Shock tubes provide well-controlled high-temperature and high-pressure conditions for chemical kinetics studies, yet the region behind the reflected shock is seldom perfectly homogeneous. Axial and radial gradients arise from shock formation, attenuation, and the interaction of the reflected shock wave with the boundary layer, and these variations influence chemical kinetic measurements such as ignition delay time. The present study combines experimental diagnostics and numerical simulations to quantify these gradients in a single-diaphragm shock tube. A coupled RANS–LES framework implemented in CONVERGE CFD incorporates realistic diaphragm opening profiles and is validated using pressure histories and shock velocity profiles for argon, nitrogen, and carbon dioxide. The results show that incident shock attenuation strongly influences the thermodynamic state of the reflected-shocked region, with test gas-dependent differences: a nearly uniform core with modest axial gradients is maintained in argon, whereas substantial axial gradients due to reflected-shock/boundary-layer interactions is seen in nitrogen and carbon dioxide. The analysis provides a foundation for quantifying test-gas homogeneity in shock-tube experiments and potential extrapolation to improving interpretation of ignition data acquired under non-ideal flow conditions.}





\maketitle
\section{Introduction}

Shock tubes are widely used canonical facilities for combustion, chemical kinetics, and high-speed gas dynamics studies~\cite{Gu_2020,Bhaskaran_2002,Reynier_2016}, with emerging applications in material science~\cite{ana@2022,Maity_2019} and biomedical engineering~\cite{Subburaj_2017}. A typical shock tube consists of a high-pressure driver section and a low-pressure driven section separated by a diaphragm. Upon diaphragm rupture, a normal incident shock wave (ISW) forms and compresses the test gas from its initial state (state 1) to the shocked state 2. After reflecting from the end wall, the reflected shock wave (RSW) further compresses the gas to state 5, where ignition delay times (IDTs) are typically measured. Under ideal assumptions, region 5 is thermodynamically homogeneous, thereby providing uniform conditions for IDT measurements. However, several recent studies have demonstrated that significant spatial and temporal inhomogeneities persist in region 5 and can substantially influence IDTs~\cite{grogan2017regimes, lipkowicz2019analysis, yamashita2012visualization}. These non-ideal conditions arise from multiple mechanisms intrinsic to shock-tube operation, as outlined below:

\begin{figure*}
    \centering
    \includegraphics[width=0.95\textwidth]{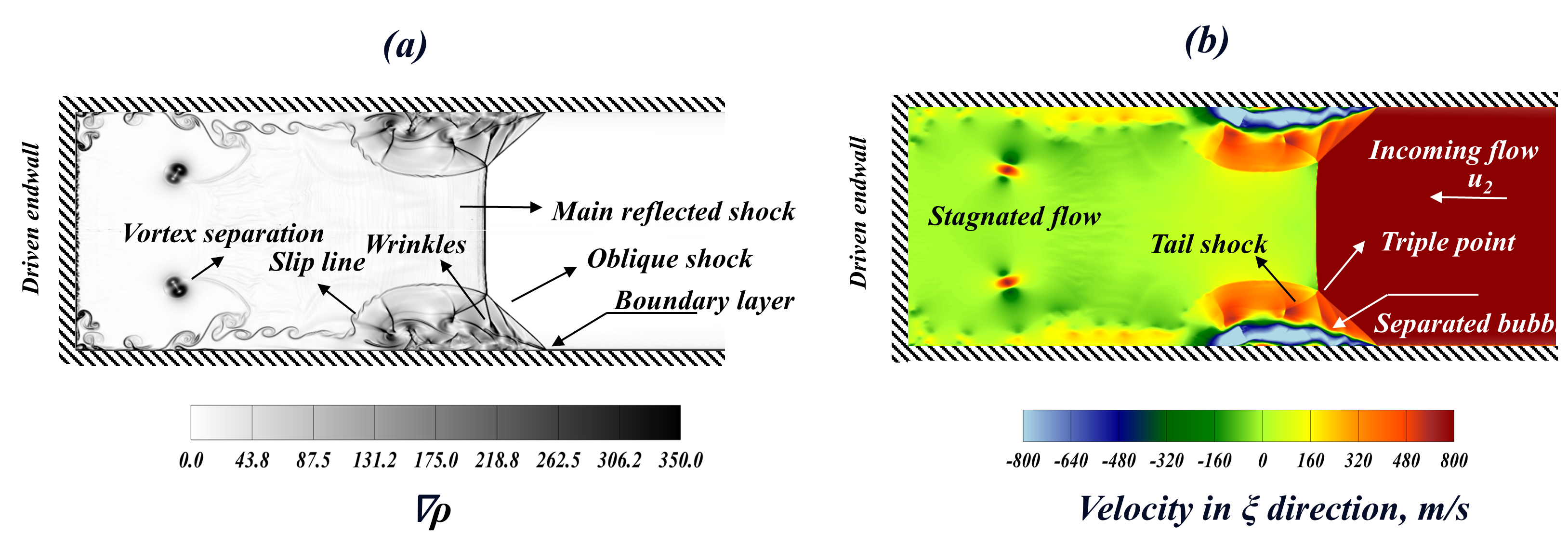}
    \caption{Typical flow field resulting from reflected shock bifurcation near the shock-tube end wall. Contours of (a) density gradient magnitude ($\nabla\rho$) and (b) axial ($\xi$) velocity field are shown. Positive velocity indicates flow towards the end wall, whereas negative velocity implies flow reversal.}
    \label{fig:intro_shock_bifurcation}
\end{figure*}

\begin{itemize}
    \item \textbf{Finite diaphragm opening time} leads to a prolonged acceleration phase of the ISW, leading to thermodynamic gradients behind the shock front~\cite{white1958influence,rothkopf1974diaphragm,kashif2025flow}.

    \item \textbf{Boundary layer (BL) behind the ISW} attenuates the shock wave~\cite{mirels1957attenuation} and further contributes to axial gradients in region 2.

    \item \textbf{RSW propagates through a non-uniform region 2 and interacts with BL} triggering flow instabilities~\cite{grogan2017regimes,petersen2001nonideal}, that introduces axial and radial thermodynamic gradients in region 5.
    
\end{itemize}

These phenomena, particularly the interaction between RSW and BL, strongly influence the resulting thermodynamic gradients and the homogeneity of region 5. The nature and extent of their interaction may induce mild flow separation or, in more severe cases, lead to a shock bifurcation that substantially reduces the homogeneous core. A theoretical framework for predicting the onset of bifurcation and the associated shock structure was first reported by Mark~\cite{mark1958interaction}. He proposed that bifurcation occurs when the stagnation pressure in the BL ($P_{0,\mathrm{BL}}$) becomes insufficient to overcome the pressure behind the RSW ($P_5$). Under these conditions, the near-wall fluid, having substantially lower momentum than the core flow, does not transmit through the RSW and is instead deflected upstream, forming a separated flow region adjacent to the wall. The separation bubble formation is depicted in Figure~\ref{fig:intro_shock_bifurcation}, that illustrates a typical flow field associated with shock bifurcation near the shock tube end wall. The density gradient (\(\nabla\rho\)) in Fig.~\ref{fig:intro_shock_bifurcation}a highlights key features and discontinuities in the flow field, while Fig.~\ref{fig:intro_shock_bifurcation}b presents axial ($\xi$) velocity contours with positive velocity indicating flow towards the end wall. The bifurcated-shock system consisting of a main RSW, an oblique shock that redirects BL flow, and a tail shock to adjust the pressure across the separation region is also shown in the figure. These shocks intersect at a triple point, from which a slip line emerges, separating fluid processed by the bifurcated shock system from the gas behind the main RSW. Entropy and velocity mismatch across the slip line lead to instabilities and vortex roll-up as the RSW propagates upstream. As more boundary-layer fluid is entrained into the separation bubble, the bifurcated structure grows in height and complexity. High-resolution simulations~\cite{ma2025numerical} show that the recirculation region evolves from an initial ``crooked-earthworm'' shape to a more stable strip-like pattern at higher shock Mach numbers. Within this structure, localized pockets of elevated pressure and temperature emerge due to nonlinear wave interactions.

\begin{table*}[t]
\centering
\renewcommand{\arraystretch}{0.95}
\resizebox{\textwidth}{!}{%
\begin{tabular}{c c c c c c c c c}
\hline
Notation & P$_1$ (mbar) & Driven gas & M$_{s,\text{end wall}}$ & T$_2$ (K) & P$_2$ (bar) & T$_5$ (K) & P$_5$ (bar) & ISW attenuation (\%/m) \\
\hline
\multirow{4}{*}{A1--A4} & 133.3 & \multirow{4}{*}{Ar}     & 3.14 & 1156 & 1.6 & 2317 & 7.1 & 1.08 \\
                        & 266.6 &                         & 2.67 & 907.6 & 2.3 & 1719 & 9.3 & 0.96 \\
                        & 533.3 &                         & 2.27 &  721 & 3.3 & 1271 & 11.7& 0.71 \\
                        & 799.9 &                         & 2.04 &  628 & 3.9 & 1051 & 12.7& 0.54 \\
\hline
\multirow{4}{*}{N1--N4} & 133.3 & \multirow{4}{*}{N$_2$}  & 3.05 &  793 & 1.4 & 1335 &  7.2& 0.81 \\
                        & 266.6 &                         & 2.57 &  645 & 2.1 & 1035 &  8.8& 0.76 \\
                        & 533.3 &                         & 2.16 &  537 & 2.84&  812 & 10.45& 0.67 \\
                        & 799.9 &                         & 1.98 &  490 & 3.5 &  713 & 11.5 & 0.48 \\
\hline
\multirow{3}{*}{C1--C3} & 133.3 & \multirow{3}{*}{CO$_2$} & 3.29 &  673 & 1.65& 1030 & 10.5 & 0.66 \\
                        & 266.6 &                         & 2.77 &  568 & 2.3 &  830 & 12.3 & 0.65 \\
                        & 533.3 &                         & 2.32 &  487 & 3.2 &  675 & 13.8 & 0.62 \\
\hline
\end{tabular}}
\caption{Summary of conditions investigated in this study. Experiments are conducted only at $P_1$ = 133.3 and 533.3 mbar.}
\label{tab:exp_conditions}
\end{table*}

Several studies have investigated shock bifurcation dynamics and their implications for flow structure and ignition behavior in shock tubes. Analytical expressions for predicting the onset of bifurcation can be found in \cite{mark1958interaction} and are summarized in section S4 in supplementary material. This framework was later refined by Davies and Wilson~\cite{davies1969influence} and Matsuo et al.~\cite{matsuo1974interaction}. Weber et al.~\cite{weber1995numerical} visualized vortex structures, reattachment shocks, and expansion fans within the bifurcated shock zone but did not consider pre-reflection gradients caused by diaphragm dynamics or ISW attenuation. Grogan and Ihme~\cite{grogan2017regimes} classified RSW–BL interactions, highlighting their influence on IDTs, though their model assumed a simplified flow over a flat plate and did not resolve thermodynamic gradients. Large-Eddy Simulations (LES) studies by Lipkowicz et al.~\cite{lipkowicz2019analysis} showed remote ignition induced due to temperature gradients but their simulations were restricted to the driven-section end and similarly neglected the coupled effects of diaphragm opening and incident-shock attenuation. Direct Numerical Simulations (DNS) revealed unsteady bifurcation growth, yet did not provide a systematic quantification of axial thermodynamic gradients~\cite{ma2025numerical}. Experimental and numerical work by Yamashita et al.~\cite{yamashita2012visualization} suggested localized ignition zones near the contact surface due to upstream temperature elevations. Collectively, these studies provide valuable insight into bifurcation physics and ignition behavior but either adopt simplified flow representations, such as flat-plate BL, or neglect ISW formation and propagation phase. Consequently, they do not capture the fully coupled influence of diaphragm dynamics, incident-shock attenuation, and reflected-shock evolution on the thermodynamic structure of region 5.

Objectives: (i) validated RANS–LES framework; (ii) quantification of region‑2 and region‑5 axial gradients; (iii) characterization of bifurcation regimes and RSW velocity evolution; (iv) provision of compact correlations for design and diagnostics.

The present study addresses these gaps by tracking the full evolution of shock tube flow, from the diaphragm rupture to end wall dynamics, to resolve the unsteady gas-dynamic features that result in axial and radial gradients. Realistic diaphragm opening profiles obtained from imaging are incorporated into the Reynolds averaged Navier-Stokes (RANS) simulations to accurately capture ISW formation and attenuation. After the ISW reaches the end wall, the RANS fields are mapped to a planar LES domain to resolve reflected-shock/boundary-layer interactions. Eleven cases are analyzed across three driven gases (argon, nitrogen, and carbon dioxide) that are chosen for their varying bifurcation tendencies. Centerline temperature and pressure fields are extracted to quantify axial gradients in regions 2 and 5. This framework incorporates spatially varying post-reflected-shock conditions to estimate where ignition is most likely to occur under non-ideal conditions. The remainder of this paper is organized as follows. Section 2 outlines the experimental and numerical methodology. Section 3 presents the results, starting with shock formation and gradients in region 2, followed by bifurcation and thermodynamic inhomogeneities in region 5. Finally, the conclusions of the study are summarized in Section 5.

\begin{figure*}[t]
    \centering
    \includegraphics[width=0.9\textwidth]{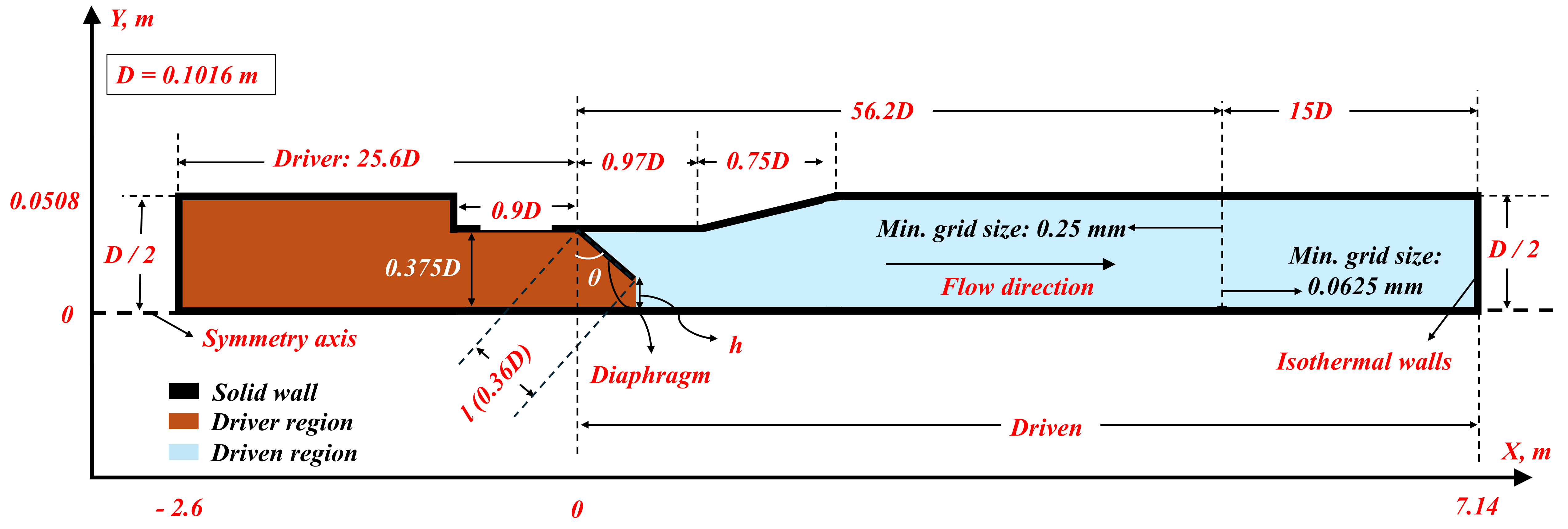}
    \caption{Schematic of the numerical domain used in the CFD simulations. In the case of RANS, a minimum grid size of 0.25\,mm is used, while a minimum cell size of 0.0625\,mm for the final 1.5\,m of the driven section is used for LES.}
    \label{fig:numerical_domain}
\end{figure*}

\section{Methodology}
\subsection{Experimental Methodology}

Experiments were conducted in the High-Pressure Shock Tube (HPST) at KAUST, featuring a 2.6\,m driver and a 7.16\,m driven section (inner diameter: 101.6\,mm). The facility configuration, diagnostics, and uncertainty quantification have been described in detail in prior studies~\cite{kashif2023effect,figueroa2023dual}. The final 0.54\,m of the driven section includes an optical section with a sapphire end wall for visualizing the diaphragm opening process. ISW velocities were measured using six PCB pressure transducers (Model: 112B05, PCB Piezotronics, USA) mounted along the driven section. The experimentally measured shock velocities were used to compute thermodynamic properties behind both the ISW and RSW using ideal shock relations. In addition to the velocity measurements, pressure traces were recorded at a fixed location of 2\,cm from the end wall to validate the numerical simulations across all test cases. Three driven gases - Ar, N$_2$, and CO$_2$ - were tested at two initial pressures (P$_{1}$ = 133.3 and 533.3 mbar) each, as summarized in Table~\ref{tab:exp_conditions}. Additional conditions at initial pressures of 233.3 and 799.9 mbar were examined numerically as described in the next section. The purity of all the gases used for the experiments was 99.999\%.

\subsection{Numerical RANS-LES framework}

ISW formation and attenuation was simulated using RANS with the $k$–$\omega$ SST turbulence model in CONVERGE v3.1\citep{converge2024}. The numerical domain replicated a 2-D planar geometry of the HPST as illustrated in Fig.~\ref{fig:numerical_domain}. Realistic diaphragm opening dynamics was incorporated via a rotating petal model as detailed in~\cite{kashif2025flow}. Driver/driven initial conditions matched the conditions used in experiments, and all walls were maintained at 294.6\,K in the simulations. Grid resolution began at 2\,mm and was refined to 0.25\,mm using adaptive mesh refinement (AMR) in regions with high gradients. Heat transfer was modeled using the Han–Reitz correlation. This RANS shock-tube framework, previously validated for predicting ISW acceleration and attenuation~\cite{kashif2025flow,kashif2025double}, is used to provide initial conditions for the LES calculations after the ISW reached the driven end wall. 

Planar 2-D LES was employed to capture the inherently unsteady reflected-shock bifurcation and the associated slip-line dynamics as RANS simulations dampen these motions and smear the triple-point structure. The objective here was to quantify axial non-uniformities with resolved unsteadiness; therefore, claims were limited to planar measures, and circumferential variability was outside the present scope. The LES framework inherited the full thermodynamic and velocity field from the RANS simulation, including the BL and axial gradients established in region 2. The LES mesh used a base cell size of 2\,mm and five levels of AMR was applied in the final 1.5\,m of the driven section to resolve the bifurcation and vortex structures near the end wall. This refinement yielded a minimum cell size of 62.5\,$\mu$m. In the remaining domain, the mesh was refined to 0.25\,mm. A separate grid independence study (see S2 in supplementary material) was conducted to confirm numerical convergence. A sigma subgrid-scale model was employed for LES turbulence closure. Wall functions were used to model near-wall velocity and temperature. Wall heat transfer was modeled using the same Han-Reitz correlation, similar to the RANS simulations. Spatial discretization was performed using a third-order monotonic upstream-centered scheme for conservation laws (MUSCL), and time integration was handled using an implicit second-order method. The LES simulations were run until $\approx$1800\,$\mu$s after ISW reflection, providing sufficient temporal scales to capture the evolution of shock bifurcation, vortex structures, and axial thermodynamic variations in region 5.

The accuracy of the CFD framework was assessed by comparing RANS-predicted ISW velocities and LES-predicted pressure histories near the end wall with the experimental data across six test cases (see section S1 in supplementary material). The RANS simulations captured the shock attenuation trends in the final section of the tube, while LES predictions closely matched the pressure rise and fluctuation patterns following shock reflection. Deviations, particularly in oscillation amplitudes, are attributed to the 2-D assumption of the domain. While 3-D LES can capture azimuthal variations, the associated computational cost is prohibitive for the spatial and temporal scales considered here. State-of-the-art 3-D RSW-BL simulations already require on the order of $\mathcal{O}(10^6)$ core-hours for substantially smaller domains (e.g., Lipkowicz et al.\cite{lipkowicz2019analysis,lipkowicz2021numerical}). Therefore, the present study is limited to a planar LES approach to resolve the unsteady RSW-BL physics that capture the axial gradients of interest.

\section{Results and discussion}

\begin{figure*}
    \centering
    \includegraphics[width=0.95\textwidth]{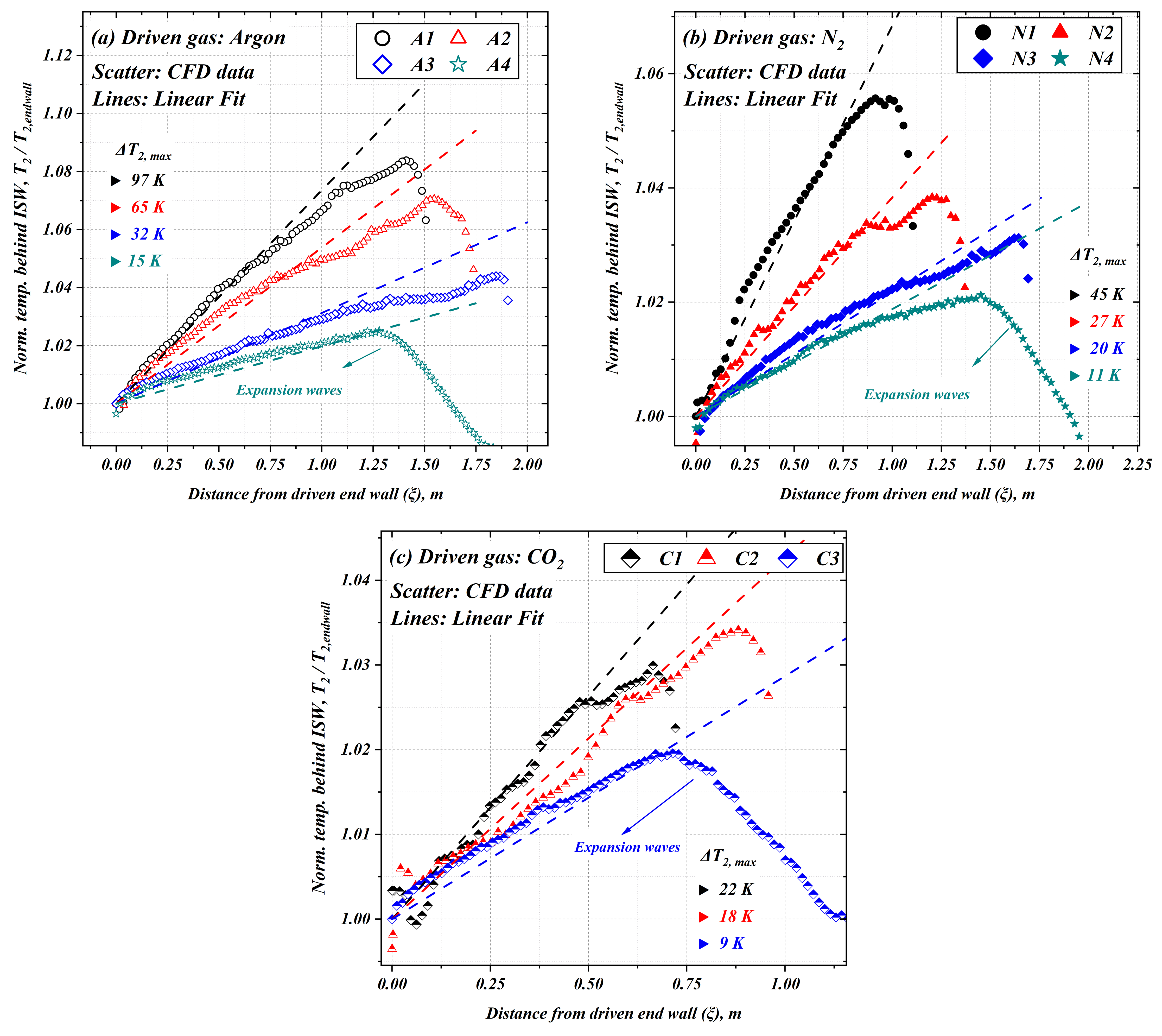}
    \caption{Numerically obtained normalized temperature values plotted against the absolute distance from the end wall, when the ISW reaches the shock tube end. Dashed lines indicate the best linear fit for the scatter points.} .
    \label{fig:axial_variation_vs_distance_temperature}
\end{figure*}

\subsection{Thermodynamic Gradients Behind ISW}

Early studies have shown that the ISW undergoes an initial acceleration phase, reaching its peak velocity as the diaphragm ruptures and the aperture opens fully ~\cite{white1958influence,rothkopf1974diaphragm} (see Fig. S1). Subsequently, BL growth and viscous dissipation results in the deceleration of ISW as it approaches the end wall. The driven gas processed initially during the ISW acceleration phase attains lower post-shock conditions due to weaker compression compared to the gas processed by the ISW at peak velocity and maximum strength. As the ISW attenuates under the influence of BL, the gas undergoes progressively weaker compression. Therefore, the varying strength of ISW along the length of the shock tube leads to an axial thermodynamic stratification in region 2. Figure S3 presents the centerline normalized temperature and pressure at four locations along the tube length (10, 25, 40, and 70\,X/D) for selected cases. Here 'X' represents the axial location along the tube and 'D' is the diameter of the shock tube. The normalization parameters ($T_{2,\text{end wall}}$ and $P_{2,\text{end wall}}$) are obtained by applying ideal shock relations for a given ISW velocity at the driven end wall. The abscissa in Figure S4 represents the normalized shocked gas length, where 0 denotes the contact surface location and 100 denotes the location of the incident shock front. During the early stages associated with the acceleration phase of the ISW (e.g., 10\,X/D), large thermodynamic gradients are observed. These gradients gradually diminish close to 25–40\,X/D, where the ISW reaches peak velocity. By 70\,X/D (i.e., when the ISW reaches the end wall), a clear linear stratification emerges, with temperature and pressure increasing upstream of the end wall. For the CO$_2$ case (C2), reflected expansion waves from the driver end influences the flow near the contact surface, reducing local pressure and temperature.

Holbeche and Spence~\cite{holbeche1964theoretical} measured a post-shock temperature difference of up to $200$~K between the early- and late-processed gas. Satchell et~al.~\cite{satchell2022flow} confirmed this numerically, although their ISW attenuation was imposed by artificially varying the driver-gas temperature. A systematic quantification of the gradients across driven gases, and its relation to attenuation and the shock Mach number $M_s$, has not been attempted before. To generalize these trends, Fig.~\ref{fig:axial_variation_vs_distance_temperature} shows normalized temperature as a function of absolute distance from the end wall when the ISW reaches the end wall. The scatter points denote numerically predicted data while the dashed lines show the best linear fit given by Eq.~\eqref{eq:Linear_fit_T2}. 
\begin{equation}
\begin{aligned}
T_2(\xi) &= T_{2,\text{end wall}} + \mathbb{F}\,\xi, \\
\mathbb{F} &= \kappa\, \cdot M_{s,\text{end wall}}^{\alpha}\, \cdot \mathrm{Attn}^{\beta}
\end{aligned}
\label{eq:Linear_fit_T2}
\end{equation}

Here, $\xi$ is the distance from the driven end wall (m); $\mathrm{Attn}$ denotes the mean shock-velocity decay rate (\%/m) evaluated over the last $4$~m of the driven section. The maximum temperature difference $(\Delta T_{2,max})$ is indicated in the plots shown in Fig.~\ref{fig:axial_variation_vs_distance_temperature}. Ar exhibits the largest temperature difference, consistent with larger ISW attenuation as compared to the N$_2$ and CO$_2$ cases. Expansion waves catch up with the shocked-gas slug in cases A4, N4, and C3, resulting in a temperature drop. To maintain a universal correlation, data points influenced by expansion waves are excluded from the fits. The corresponding ranges of $M_{s,\text{end wall}}$ and attenuation used in Eq.~\eqref{eq:Linear_fit_T2}, along with the fit coefficients $(\kappa,\alpha,\beta)$, are listed in Table~\ref{tab:kappa_alpha_beta_grouped}.

\begin{table}[t]
\centering
\resizebox{\columnwidth}{!}{%
\begin{tabular}{lccccc}
\hline
\textbf{Cases} & \textbf{$M_{s,\mathrm{end wall}}$} & \textbf{Attn (\%/m)} & \textbf{$\kappa$} & \textbf{$\alpha$} & \textbf{$\beta$} \\
\hline
A1--A4 & \(2.04\text{--}3.14\) & \(0.54\text{--}1.08\) & 4.9 & 2.5 & 1.23 \\
N1--N4 & \(1.98\text{--}3.05\) & \(0.48\text{--}0.81\) & 1.38 & 4.82 & -0.63 \\
C1--C3 & \(2.32\text{--}3.29\) & \(0.62\text{--}0.66\) & 35.4 & 1.74 & 5.14 \\
\hline
\end{tabular}}
\caption{$M_{s,\mathrm{end wall}}$, attenuation (Attn), and fit parameters $(\kappa,\alpha,\beta)$. for the investigated conditions.}
\label{tab:kappa_alpha_beta_grouped}
\end{table}

\begin{figure*}
    \centering
    \includegraphics[width=0.99\textwidth]{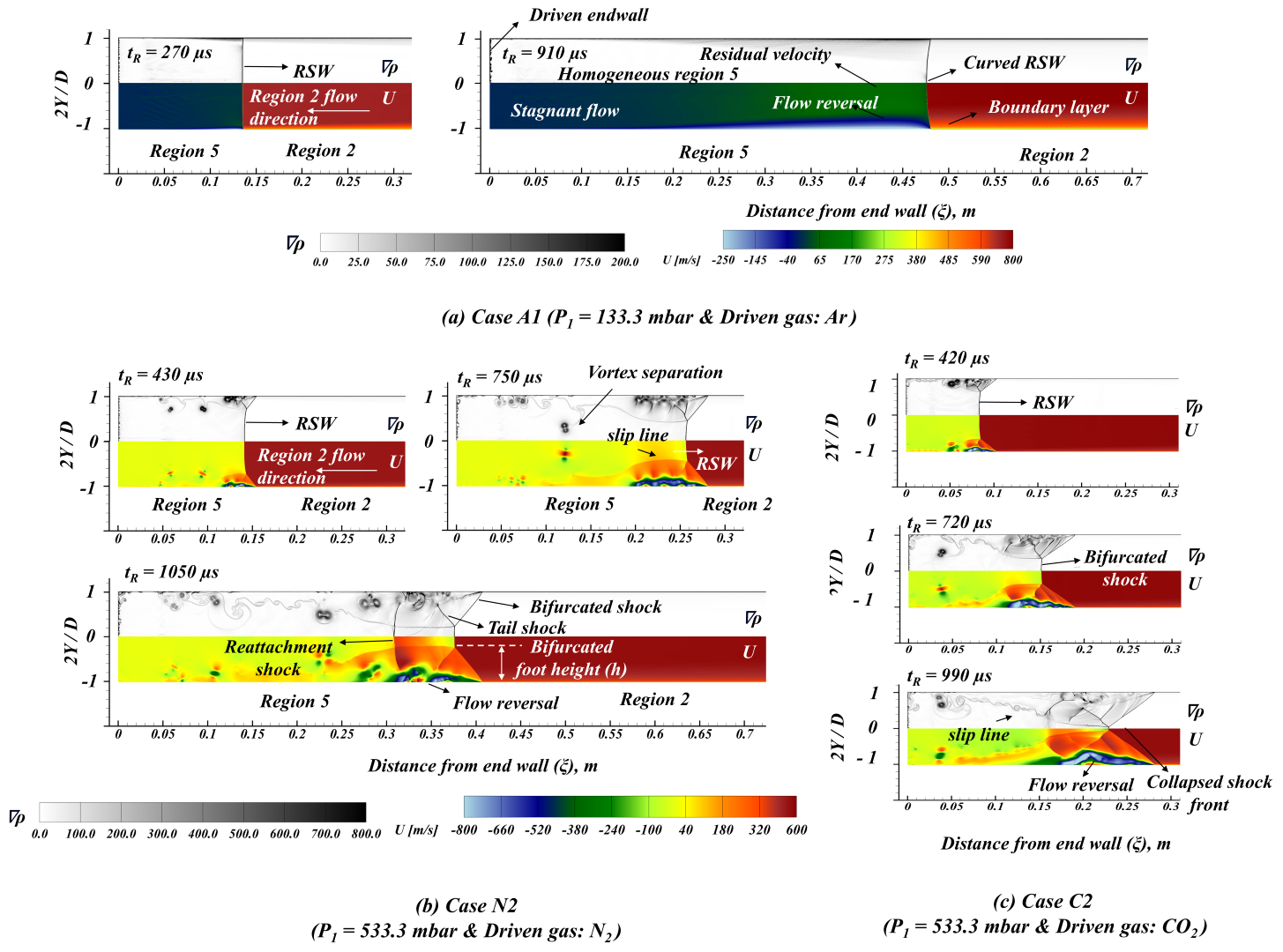}
    \caption{Flow evolution in region 5 for A1, N2, and C2. Each contour shows density gradient magnitude ($\nabla \rho$, grayscale) and axial velocity ($U$, color) at three time instances after ISW reflection. Argon shows no shock bifurcation but exhibits weak flow reversal and a curved RSW. Nitrogen and CO$_2$ display pronounced shock bifurcation, slip-line growth, and vortex formation. The CO$_2$ case exhibits a collapse of the planar shock due to stronger RSW-BL interaction at later stages.} .
    \label{fig:RSW_BL_interaction}
\end{figure*}

\subsection{RSW Bifurcation Dynamics}

Mark’s criterion~\cite{mark1958interaction} states that shock bifurcation occurs when the stagnation pressure in BL ($P_{0,\mathrm{BL}}$) becomes lesser than the post-reflected-shock pressure, i.e., $P_{0,\mathrm{BL}} / P_5 < 1$. Figure~S5 presents the computed ratio $P_{0,\mathrm{BL}} / P_5$ as a function of the $M_s$ for the three driven gases being investigated. The results show that the N$_2$ and CO$_2$ cases are prone to bifurcation for $M_s \geq 1.5$, while argon remains largely unaffected across the examined range. The minor deviation near $M_s \approx 2$ in argon appears to be an artifact of Mark's model, as both earlier studies \cite{lipkowicz2021numerical,miller1975incident} and current LES results show no evidence of bifurcation in argon. Figure~\ref{fig:RSW_BL_interaction} shows the development of RSW-BL interactions for representative cases for argon (A1), nitrogen (N2), and carbon dioxide (C2) as driven gases. Density gradient magnitude ($\nabla \rho$) and axial velocity ($U$) contours are shown at different time stamps following ISW reflection. The flow evolution in region 5, with particular emphasis on bifurcation structures and their implications on thermodynamic non-uniformity, is described below.

\textbf{Incipient Separation in Ar:} In the cases A1 - A4, the RSW-BL interaction does not result in shock bifurcation. Instead, an incipient separation occurs near the wall, accompanied by weak circulation zones and a gradual distortion of the RSW front. Figure~\ref{fig:RSW_BL_interaction}a shows contours of density gradient magnitude and axial velocity ($U$) for case A1 at various time stamps. Positive values of $U$ indicate gas motion towards the end wall. In the initial stages post-ISW reflection ($t_R = 270$\,$\mu$s), the gas behind the RSW is nearly stagnant, forming a quasi-steady region 5. As the RSW propagates upstream, two key flow features become apparent: (i) the development of wall-attached circulation regions due to flow reversal of BL fluid, and (ii) a gradual curvature of the RSW front, particularly closer to the walls ($t_R = 910$\,$\mu$s). These features arise due to the inability of low-momentum BL fluid to transmit through the RSW, leading to localized flow reversal and upstream entrainment. The curvature of the RSW can be attributed to the faster propagation of the shock front through a turbulent BL, as reported by Lipkowicz et al.~\cite{lipkowicz2019analysis}. The gas behind the reflected shock does not completely stagnate but retains a residual velocity towards the end wall. This motion compresses the gas ahead, leading to a slow and approximately linear rise in end-wall pressure (see Fig. S2), commonly referred to as dP/dt. Despite the continuous entrainment and the flow reversal in the shear layer, the fluid near the wall does not have sufficient momentum to trigger shock bifurcation. The progressive growth of the shear layer effectively forms a Laval nozzle–like constriction for the region-2 gas near the core. This constriction leads to expansion of the gas behind the RSW, as observed in Fig.~\ref{fig:RSW_BL_interaction}a at $t_R = 910$\,$\mu$s. Collectively, these flow phenomena introduce significant thermodynamic gradients behind the RSW.

\begin{figure*}[t]
    \centering
    \includegraphics[width=0.8\textwidth]{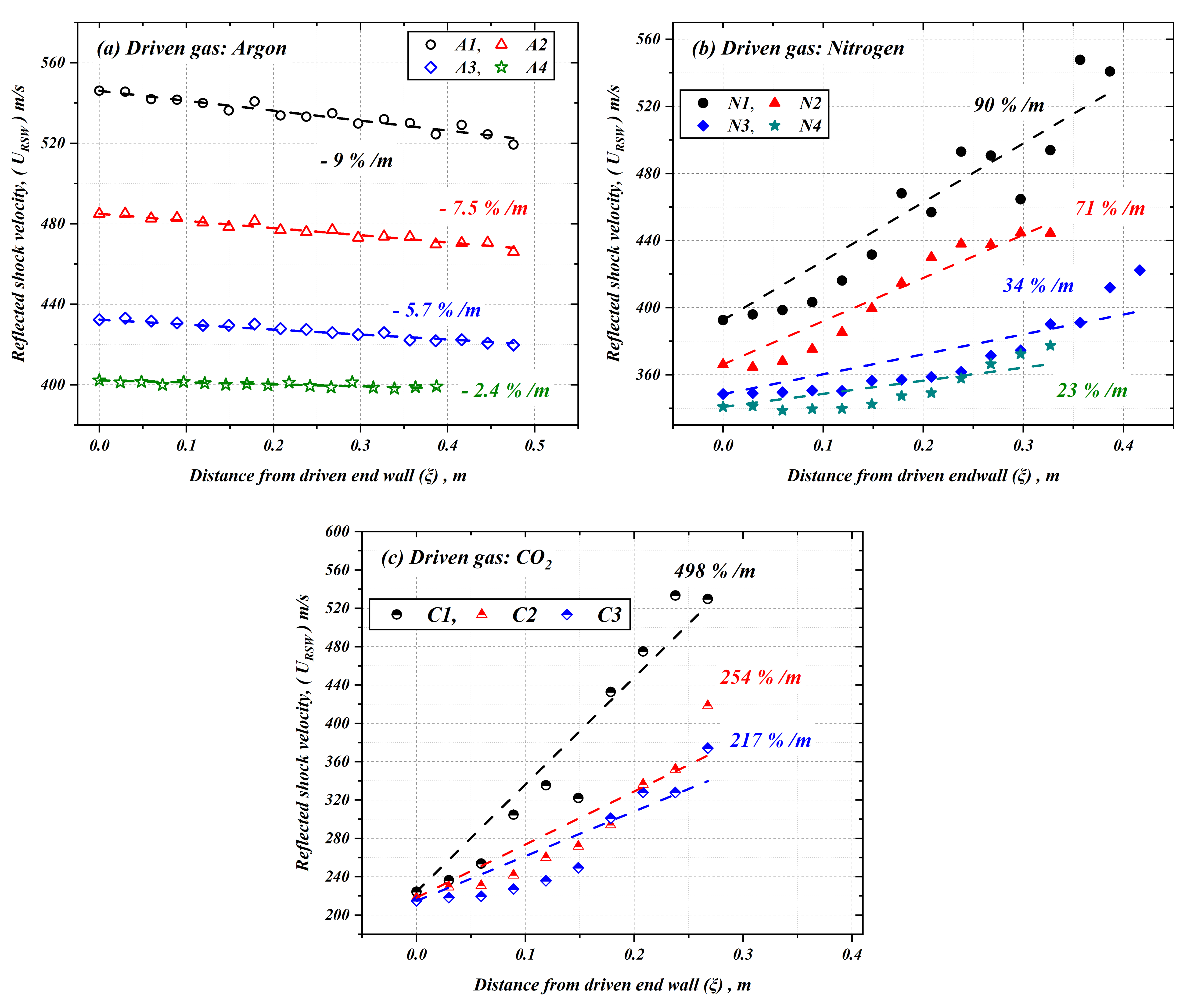}
    \caption{Centerline velocity of the RSW as a function of axial distance from the driven end wall for (a) argon (A1–A4), (b) nitrogen (N1–N4), and (c) carbon dioxide (C1–C3).}
    \label{fig:rsw_velocity}
\end{figure*}

\textbf{RSW Bifurcation and Shear Layer Evolution in N$_2$ and CO$_2$:} In contrast to the argon cases, both nitrogen (N1 - N4) and carbon dioxide (C1 - C3) exhibit pronounced shock bifurcation following RSW-BL interaction. Figures~\ref{fig:RSW_BL_interaction}b and c show the flow features in N2 and C2 cases, respectively. Contours of density gradient magnitude ($\nabla \rho$) and axial velocity ($U$) highlight the evolution of inviscid and shear-layer-dominated features. In both cases, the low-momentum BL fluid is unable to transmit through the RSW, leading to the formation of a triple-point structure composed of the main curved RSW, an oblique shock, and a foot (or tail) shock. This deflects the near-wall fluid into a shear layer that is entrained upstream, while the inviscid core stagnates behind the RSW. By $t_R = 750\,\mu$s in N$_2$ and $t_R = 720\,\mu$s in CO$_2$, a well-defined slip line emerges behind the RSW. 
As the RSW propagates upstream, the bifurcated foot height ($h$) grows almost linearly with the distance from the end wall in both cases, reaching up to 40\,mm before the RSW encounters the contact surface (Fig. S6a). The bifurcation angle ($\alpha$) remains unchanged throughout this evolution (Fig. S6b), consistent with observations by Matsuo et al.~\cite{matsuo1974interaction}. The shear layer rolls up to form vortices towards the core as time progresses, which grow over time and enhance mixing. By $t_R = 1050\,\mu$s in N$_2$, the inviscid core is constricted to a nozzle-like geometry, that causes flow acceleration and local expansion. The pressure of the shocked gas drops and a reattachment shock is formed that recompresses the flow. This phenomenon, previously reported by Weber et al.~\cite{weber1995numerical} and Lipkowicz et al.~\cite{lipkowicz2019analysis}, compresses the core gas and re-establishes near-stagnation conditions. The resulting expansion–compression sequence is clearly visible in the velocity contours.

CO$_2$ exhibits similar trends but with more severe bifurcation. The onset occurs earlier ($t_R = 420\,\mu$s), and the bifurcated structure is more vertically extended than in N$_2$, indicating a larger momentum deficit in the BL fluid. By $t_R = 990\,\mu$s, the slip line becomes unstable, and the RSW front shows signs of collapse near the centerline. Flow reversal near the wall is more intense and reaches further vertically, displacing the inviscid region and narrowing the core passage. Vortex roll-up is also more pronounced, leading to enhanced radial mixing. In both gases, the continued growth of the bifurcated structure reduces the extent of the homogeneous core behind the RSW. The combined effects of strong shear-layer activity and axial/radial gradients significantly alter the thermodynamic field in region 5. These effects are more pronounced in CO$_2$, likely due to its higher post-shock pressures, stronger boundary-layer interaction, and lower specific heat ratio.

\begin{figure}
    \centering
    \includegraphics[width=0.5\textwidth]{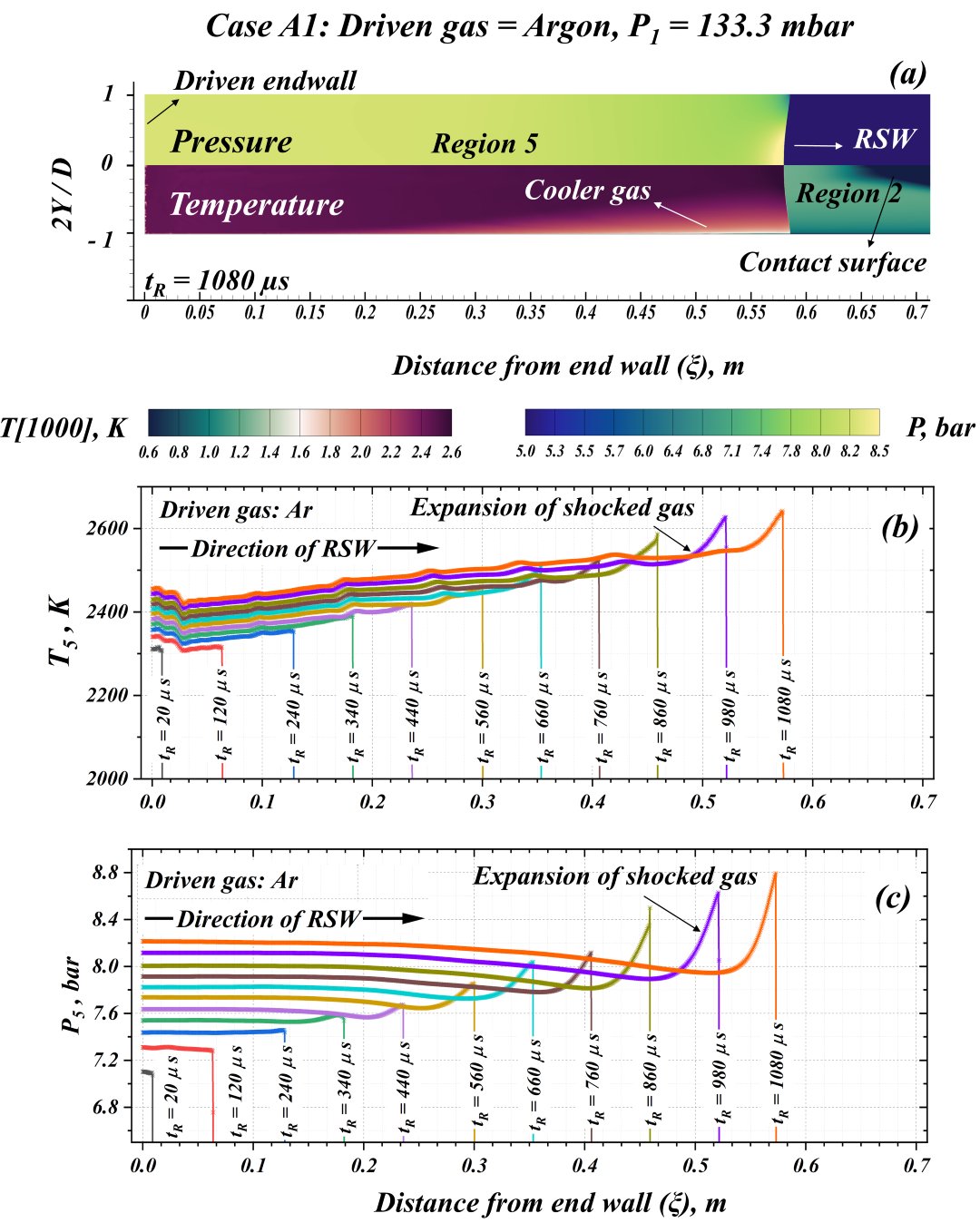}
    \caption{(a) Pressure and temperature contours at $t_R = 1080\,\mu$s for case A1. (b) Centerline temperature and (c) centerline pressure, as a function of distance from the driven end wall at different times after shock reflection.}
    \label{fig:axial_variations__region_5_argon}
\end{figure}

\subsection{Velocity Variation of RSW}
RSW is often assumed to propagate at constant speed by neglecting the effects of BL interactions and other non-ideal processes. As shown in the previous section, shock bifurcation effects can significantly alter RSW propagation.  Therefore, it is important to characterize evolution of RSW velocity, as it directly governs the spatial uniformity of $P_5$ and $T_5$. Figure~\ref{fig:rsw_velocity} shows the centerline RSW velocity as a function of distance from the end wall for all the cases considered. Symbols denote numerically obtained data points, and the dashed lines are linear fits. The velocity is tracked only within region~5, ahead of the contact surface (CS) and prior to the arrival of expansion waves, beyond which thermodynamic analysis is no longer meaningful.

In argon (see Fig.~\ref{fig:rsw_velocity}a), the RSW shows monotonic attenuation with average deceleration rates of $2.4$–$9\,\%/\mathrm{m}$. These values are consistent with experimentally measured RSW deceleration of $3.81\,\%/\mathrm{m}$ in $n$-heptane/argon mixtures reported by Susa et~al.~\cite{susa2022flame}. The absence of bifurcation in argon leads to steady momentum loss and progressive weakening of RSW. In nitrogen, the RSW exhibits acceleration, with the velocity evolution depending on the operating conditions. In case N1, the RSW velocity increases at a rate of $\sim\!90\,\%/\mathrm{m}$. In case N2, the velocity remains nearly constant over the first $\sim\!0.1\,\mathrm{m}$ before accelerating, resulting in an overall rate of $\sim\!71\,\%/\mathrm{m}$ (see Fig.~\ref{fig:rsw_velocity}b). Similar trends are observed in the remaining nitrogen cases. This behavior is consistent with experimental observations by Matsuo et~al.~\cite{matsuo1974interaction}, who reported a two-phase evolution in RSW velocity characterized by an initial plateau followed by acceleration, driven by shear-layer dynamics and growth of the bifurcated foot. Carbon dioxide (Fig.~\ref{fig:rsw_velocity}c) exhibits the same qualitative pattern but with stronger RSW acceleration, with overall rates in the range $217$–$498\,\%/\mathrm{m}$. In all the cases showing shock bifurcation (in N$_2$ and CO$_2$), the RSW propagation typically transitions from an initial near-constant velocity to pronounced acceleration. 


\subsection{Thermodynamic gradients behind RSW}

\begin{figure*}
    \centering
    \includegraphics[width=\textwidth]{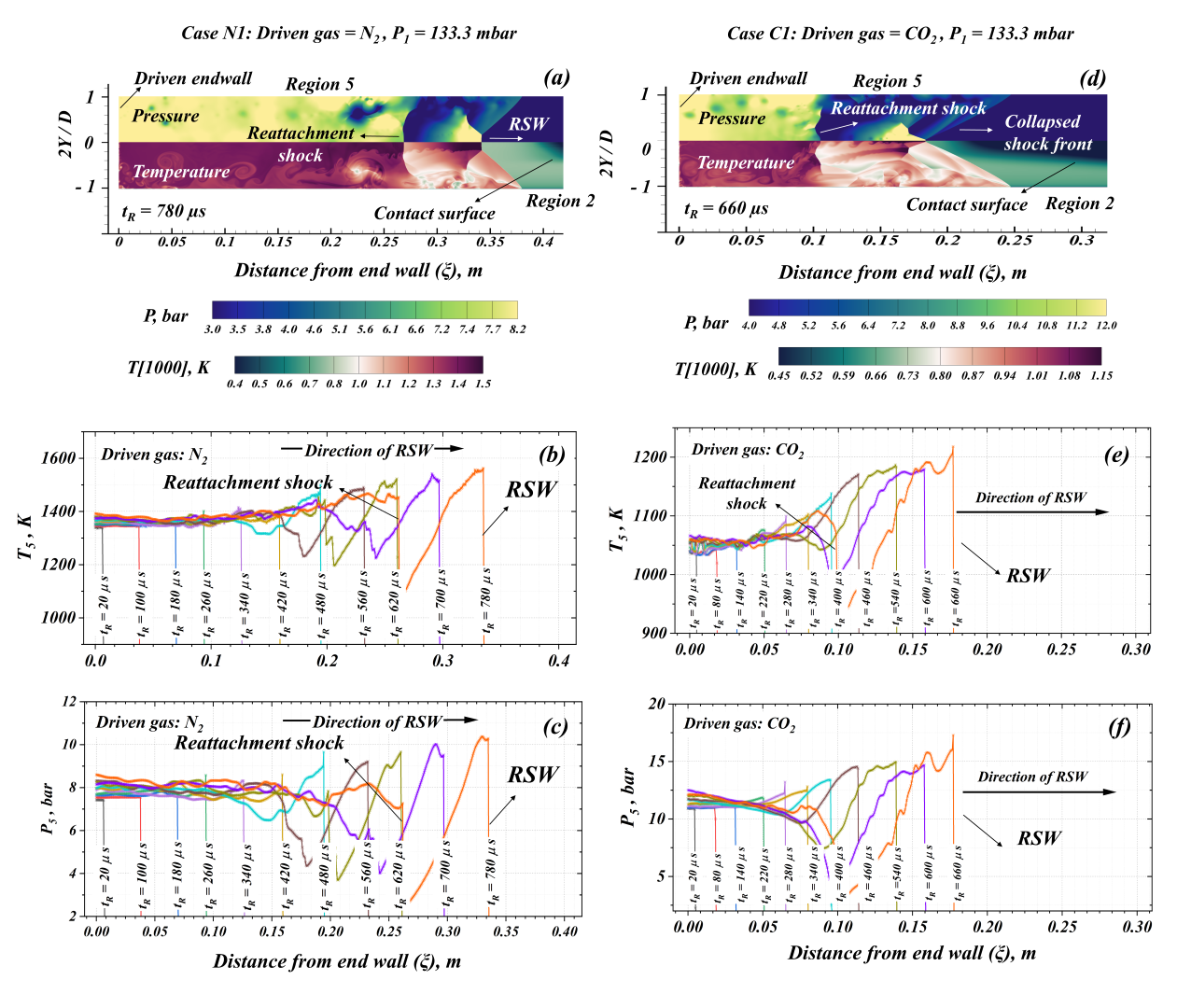}
    \caption{(a) Pressure and temperature contours at $t_R = 780\,\mu$s for case N1. (b) Centerline temperature and (c) centerline pressure, as a function of distance from the driven end wall at different times after shock reflection. (d) Pressure and temperature contours at $t_R = 660\,\mu$s for case C1. (e) Centerline temperature and (f) centerline pressure, as a function of distance from the driven end wall at different times after shock reflection.}
    \label{fig:axial_variations__region_5_n2_co2}
\end{figure*}

\begin{figure*}
    \centering
    \includegraphics[width=0.8\textwidth]{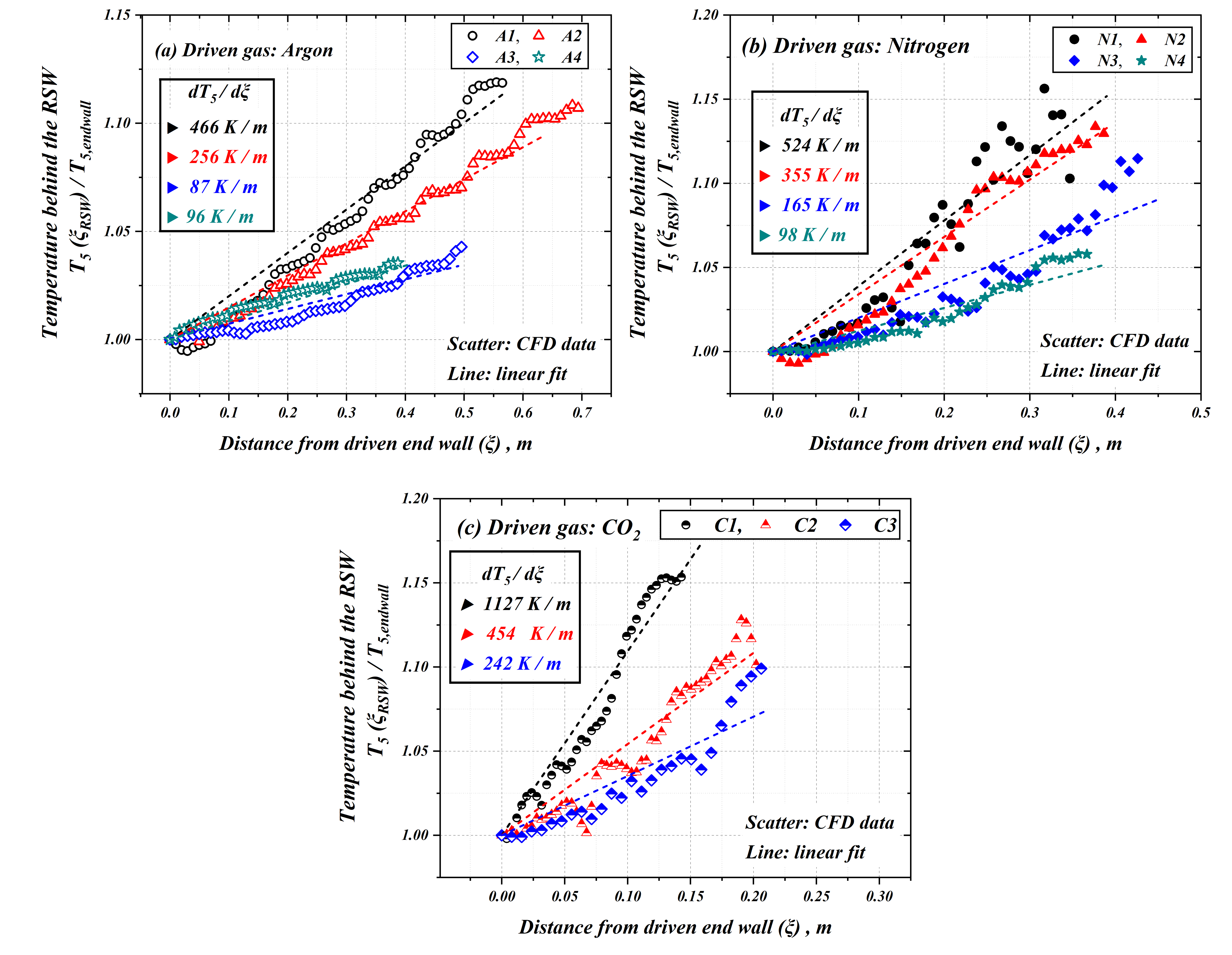}
    \caption{Normalized temperature at various locations in the shock tube at 20 $\mu$s post-RSW for cases (a) A1 - A4, (b) N1 -  N4, (c) C1 - C3.}
    \label{fig:T5_gradient}
\end{figure*}

Both radial and axial variations in thermodynamic parameters are observed in the region behind the RSW. Radial variations are primarily linked to the growth of the BL–entrained shear layer and shock bifurcation, while axial variations occur from a combination of gradients behind ISW and the evolving RSW dynamics (including velocity variation and bifurcation). 
The present analysis focuses on thermodynamic trends behind the RSW along the centerline. Although significant radial variations in parameters are expected for certain conditions examined in this study, their quantitative characterization is beyond the scope of the present study. Figures~\ref{fig:axial_variations__region_5_argon} and \ref{fig:axial_variations__region_5_n2_co2} show the spatial and temporal evolution of $T_5(\xi)$ and $P_5(\xi)$ at different time instances post-shock reflection for three representative cases: A1 (argon), N1 (nitrogen), and C1 (carbon dioxide). These cases are selected to represent different bifurcation regimes, and the trends observed here are qualitatively similar to the remaining three cases, which are discussed later. 

\textbf{Cases A1 (Argon):}  
    Figure~\ref{fig:axial_variations__region_5_argon}a presents pressure and temperature contours at $t_R=1080~\mu\mathrm{s}$ as well as  centerline $T_5$ and $P_5$ profiles at several $t_R$ instants. At the end wall ($\xi=0$), both $T_5$ and $P_5$ (see Figures~\ref{fig:axial_variations__region_5_argon}b and c) increase monotonically with $t_R$, consistent with residual-velocity effects in region~5~\cite{susa2022flame}. At upstream locations ($\xi>0$), $T_5$ increases almost linearly with $\xi$ until $t_R=240~\mu\mathrm{s}$. At later times, both $T_5$ and $P_5$ exhibit a shallow dip immediately behind the RSW because of the near-wall separation of the BL that induces mild expansion of the freshly processed gas. 

\textbf{Case N1 (nitrogen):} 
    The centerline profiles evolve more complexly in this case due to the onset of bifurcation and the formation of a reattachment shock. The pressure and temperature contours in Fig.~\ref{fig:axial_variations__region_5_n2_co2}a highlight the flow features which include the triple point, slip line, and a vortex-laden shear layer. In Figs.~\ref{fig:axial_variations__region_5_n2_co2}b and c, the $T_5(\xi)$ and $P_5(\xi)$ profiles at $t_R=480~\mu\mathrm{s}$ show a relaxation zone (drop in pressure and temperature) behind the RSW. This relaxation is attributed to the core-flow contraction as the shear layer grows, accelerating the shocked gas and reducing static pressure (local expansion). This is also visible in the pressure field shown in Fig.~\ref{fig:axial_variations__region_5_n2_co2}a. The reattachment shock is seen at $t_R=780~\mu\mathrm{s}$ that compresses the flow, resulting in the increase in pressure and temperature. Notably, the temperature immediately behind the RSW exceeds that behind the reattachment shock, indicating a stronger but short-lived initial compression, consistent with the rapid RSW acceleration observed in Fig.~\ref{fig:rsw_velocity}b. The pressure profiles exhibit a similar trend, with a temporary dip ahead of reattachment followed by recovery.

\textbf{Case C1 (Carbon Dioxide):}  
    The behavior in CO$_2$ is similar to that in N$_2$, but with more pronounced features. As shown in Figs.~\ref{fig:axial_variations__region_5_n2_co2}d-f, both temperature and pressure profiles show stronger gradients in region 5. Reattachment shocks forms earlier and the separation bubble is larger, leading to more significant variations in the centerline thermodynamic fields. There is a more severe drop in pressure and temperature in the relaxation zone ahead of reattachment shock. 

Overall, the centerline thermodynamic fields in region 5 exhibit significant axial variation, particularly in scenarios with shock bifurcation. Although some of these variations are short-lived, they can still have important implications for ignition behavior. The axial temperature rise behind the RSW is evaluated at each location $20~\mu\mathrm{s}$ after its passage and is shown in Fig.~\ref{fig:T5_gradient}. Case A1 (see Fig.~\ref{fig:T5_gradient}a) shows an approximately linear variation with a slope $\mathrm{d}T_5/\mathrm{d}\xi \approx 466~\mathrm{K\,m^{-1}}$. A similar behavior is observed for cases A2–A4 with slopes of $256$, $87$, and $96~\mathrm{K\,m^{-1}}$, respectively. Figure~\ref{fig:T5_gradient}b reports the peak normalized temperatures for the N$_2$ cases. Linear fits yield slopes in the range of $524$–$98~\mathrm{K\,m^{-1}}$. Although the fits do not capture the behavior close to the end wall, they reproduce the overall trend well. The temperatures for the CO$_2$ cases, shown in Fig.~\ref{fig:T5_gradient}c, indicate slopes of 1127, 454, and 242 $~\mathrm{K\,m^{-1}}$ for C1 - C3, respectively. 

\section{Conclusion}\label{sec13}

This study investigates thermodynamic inhomogeneities behind the ISW and RSW in a single-diaphragm shock tube using a coupled RANS-LES simulations with experimentally measured diaphragm opening profiles. The modeling framework accurately captures shock formation dynamics, with incident shock velocities showing good agreement with experiments. Axial gradients behind the ISW were found to vary significantly in different gases, with argon exhibiting the steepest temperature gradient, reaching up to 8.4\% of the end wall value ($\approx$97 K). These gradients were quantified and correlated with the end wall Mach number and shock attenuation. The reflected shock region (region 5) exhibits a strongly gas-dependent behavior: argon shows incipient separation, whereas nitrogen and carbon dioxide developed pronounced shock bifurcation, with CO$_2$ displaying more severe features that lead to significant reduction in the homogeneous core. These bifurcated structures alter the RSW velocity, which decelerates in argon but accelerates by 23 – 500\%/m in N$_2$ and CO$_2$. Associated reattachment shock and expansion–compression regions disrupt the thermodynamic uniformity of region 5, generating axial temperature gradients as high as 524 $~\mathrm{K\,m^{-1}}$ and 1127$~\mathrm{K\,m^{-1}}$ in N$_2$ and CO$_2$, respectively. Future experiments will focus on validating the key observations made in this work, such as RSW acceleration and axial gradients in temperature in region 5.

\backmatter

\bmhead{Supplementary information}

The supplementary files include videos on the valve opening measurements and flow visualization near the valve from CFD simulations.

\bmhead{Acknowledgments}

This work was sponsored by King Abdullah University of Science and Technology (KAUST) and supported by the KAUST Supercomputing Laboratory (KSL). All simulations were performed on KSL’s Shaheen III supercomputer. Convergent Science provided CONVERGE licenses and technical support for this work

\bmhead{Data availability}
Datasets generated during the current study are available from the corresponding authors on reasonable request.

\bmhead{Competing interests}
The authors declare no competing interests.

\bibliography{sn-bibliography}

@article{Gu_2020,
title = {Capabilities and limitations of existing hypersonic facilities},
journal = {Prog. Aerosp. Sci.},
volume = {113},
pages = {100607},
year = {2020},
issn = {0376-0421},
author = {Sangdi Gu and Herbert Olivier},
}

@article{Reynier_2016,
title = {Survey of high-enthalpy shock facilities in the perspective of radiation and chemical kinetics investigations},
journal = {Prog. Aerosp. Sci.},
volume = {85},
pages = {1-32},
year = {2016},
issn = {0376-0421},
author = {Philippe Reynier},
}

@article{Bhaskaran_2002,
title = {The shock tube as wave reactor for kinetic studies and material systems},
journal = {Prog. Energy Combust. Sci.},
volume = {28},
pages = {151-192},
year = {2002},
issn = {0360-1285},
author = {K.A Bhaskaran and P Roth},
}

@article{ana@2022,
title = {Shockwave impact on the stability of anatase titania nanoparticles},
journal = {Mater. Today Commun.},
volume = {32},
pages = {104031},
year = {2022},
issn = {2352-4928},
author = {Ana Luiza {Slama de Freitas} and Janardhanraj Subburaj and Juan Carlos Navarro and Hassnain Abbas Khan and Touqeer Anwar Kashif and Khaiyom Hakimov and Javier Ruiz-Martinez and Aamir Farooq},
}

@article{Maity_2019,
author = {Maity, Tarak N. and Gopinath, Nagarajan K. and Janardhanraj, S. and Biswas, Krishanu and Basu, Bikramjit},
title = {Computational and Microstructural Stability Analysis of Shock Wave Interaction with NbB2-B4C-Based Nanostructured Ceramics},
journal = {ACS Appl. Mater. Interfaces},
volume = {11},
pages = {47491-47500},
year = {2019},
}

@Article{Subburaj_2017,
author={Subburaj, Janardhanraj and Datey, Akshay and Gopalan, Jagadeesh and Chakravortty, Dipshikha},
title={Insights into the mechanism of a novel shockwave-assisted needle-free drug delivery device driven by in situ-generated oxyhydrogen mixture which provides efficient protection against mycobacterial infections},
journal={J. Biol. Eng.},
year={2017},
month={Dec},
day={12},
volume={11},
pages={48},
issn={1754-1611},
}

@article{white1958influence,
  title={Influence of diaphragm opening time on shock-tube flows},
  author={White, Donald R},
  journal={J. Fluid Mech.},
  volume={4},
  pages={585--599},
  year={1958},
  publisher={Cambridge University Press}
}

@article{rothkopf1974diaphragm,
  title={Diaphragm opening process in shock tubes},
  author={Rothkopf, EM and Low, W},
  journal={Phys. Fluids},
  volume={17},
  pages={1169--1173},
  year={1974},
  publisher={American Institute of Physics}
}

@article{miller1975incident,
  title={Incident shock-wave characteristics in air, argon, carbon dioxide, and helium in a shock tube with unheated helium driver},
  author={Miller III, Charles G and Jones, Jim J},
  journal={NASA Tech. Note D-8099 (L-10520), NASA Langley Research Center},
  year={1975}
}

@article{kashif2023effect,
  title={Effect of oxygen enrichment on methane ignition},
  author={Kashif, Touqeer Anwar and AlAbbad, Mohammed and Figueroa-Labastida, Miguel and Chatakonda, Obulesu and Kloosterman, Jeffrey and Middaugh, Joshua and Sarathy, S Mani and Farooq, Aamir},
  journal={Combust. Flame},
  volume={258},
  pages={113073},
  year={2023},
  publisher={Elsevier}
}

@article{figueroa2023dual,
  title={Dual-camera high-speed imaging of n-hexane oxidation in a high-pressure shock tube},
  author={Figueroa-Labastida, Miguel and Kashif, Touqeer Anwar and Farooq, Aamir},
  journal={Combust. Flame},
  volume={248},
  pages={112586},
  year={2023},
  publisher={Elsevier}
}

@article{mirels1957attenuation,
  title={Attenuation in a shock tube due to unsteady-boundary-layer action},
  author={Mirels, Harold},
  journal={NACA Rep. 1333, National Advisory Committee for Aeronautics},
  year={1957}
}

@article{satchell2022flow,
  title={Flow nonuniformities behind accelerating and decelerating shock waves in shock tubes},
  author={Satchell, Matthew and Di Mare, Luca and McGilvray, Matthew},
  journal={AIAA J.},
  volume={60},
  pages={1537--1548},
  year={2022},
  publisher={American Institute of Aeronautics and Astronautics}
}

@article{grogan2017regimes,
  title={Regimes describing shock boundary layer interaction and ignition in shock tubes},
  author={Grogan, Kevin P and Ihme, Matthias},
  journal={Proc. Combust. Inst.},
  volume={36},
  pages={2927--2935},
  year={2017},
  publisher={Elsevier}
}

@article{lipkowicz2019analysis,
  title={Analysis of mild ignition in a shock tube using a highly resolved {3D}-{LES} and high-order shock-capturing schemes},
  author={Lipkowicz, JT and Wlokas, I and Kempf, AM},
  journal={Shock Waves},
  volume={29},
  pages={511--521},
  year={2019},
  publisher={Springer}
}

@article{lipkowicz2021numerical,
  title={Numerical investigation of remote ignition in shock tubes},
  author={Lipkowicz, Jonathan Timo and Nativel, Damien and Cooper, Sean and Wlokas, Iren{\"a}us and Fikri, Mustapha and Petersen, Eric and Schulz, Christof and Kempf, Andreas Markus},
  journal={Flow Turbul. Combust.},
  volume={106},
  pages={471--498},
  year={2021},
  publisher={Springer}
}

@article{kashif2025flow,
  title={On the flow characteristics in the shock formation region due to the diaphragm opening process in a shock tube},
  author={Kashif, Touqeer Anwar and Subburaj, Janardhanraj and Farooq, Aamir},
  journal={Phys. Fluids},
  volume={37},
  year={2025},
  pages={066117},
  publisher={AIP Publishing}
}

@article{kashif2025double,
  title={Double-Diaphragm Induced Shock Velocity Variation and Its Effects on Shocked Gas},
  author={Kashif, Touqeer Anwar and Subburaj, Janardhanraj and Ali Khan, Md Zafar and Farooq, Aamir},
  journal={AIAA J.},
  volume={63},
  pages={5059--5072},
  year={2025},
  publisher={American Institute of Aeronautics and Astronautics}
}

@book{converge2024,
  title={CONVERGE 3.1.11},
  author={Richards, K. J. and Senecal, P. K. and Pomraning, E.},
  year={2024},
  publisher={Convergent Science, Madison, WI, USA}
}

@article{petersen2001nonideal,
  title={Nonideal effects behind reflected shock waves in a high-pressure shock tube},
  author={Petersen, Eric L and Hanson, Ronald K},
  journal={Shock Waves},
  volume={10},
  pages={405--420},
  year={2001},
  publisher={Springer}
}

@article{mark1958interaction,
  title={The interaction of a reflected shock wave with the boundary layer in a shock tube},
  author={Mark, Herman},
  journal={NACA-TM-1418, National Advisory Committee for Aeronautics},
  year={1958}
}

@article{matsuo1974interaction,
  title={The interaction of a reflected shock wave with the boundary layer in a shock tube},
  author={Matsuo, Kazuyasu and Kawagoe, Shigetoshi and Kage, Kazuyuki},
  journal={Bull. JSME},
  volume={17},
  pages={1039--1046},
  year={1974},
  publisher={The Japan Society of Mechanical Engineers}
}

@article{davies1969influence,
  title={Influence of Reflected Shock and Boundary-Layer Interaction on Shock-Tube Flows},
  author={Davies, L and Wilson, JL},
  journal={Phys. Fluids},
  volume={12},
  pages={37-43},
  year={1969},
  publisher={AIP Publishing}
}

@article{ma2025numerical,
  title={Numerical Study of Bifurcation Structures in Reflected Shock-Wave/Laminar-Boundary-Layer Interaction Within an End-Wall Tube},
  author={Ma, Zhuang and Lee, Shibo and Zhao, Yunlong and Zhang, Yang},
  journal={J. Fluids Eng.},
  volume={147},
  year={2025},
  pages={041202},
  publisher={American Society of Mechanical Engineers Digital Collection}
}

@article{weber1995numerical,
  title={The numerical simulation of shock bifurcation near the end wall of a shock tube},
  author={Weber, Yvette S and Oran, ES and Boris, JP and Anderson Jr, JD},
  journal={Phys. Fluids},
  volume={7},
  pages={2475--2488},
  year={1995},
  publisher={American Institute of Physics}
}

@article{yamashita2012visualization,
  title={Visualization study of ignition modes behind bifurcated-reflected shock waves},
  author={Yamashita, Hiroki and Kasahara, Jiro and Sugiyama, Yuta and Matsuo, Akiko},
  journal={Combust. Flame},
  volume={159},
  pages={2954--2966},
  year={2012},
  publisher={Elsevier}
}

@article{holbeche1964theoretical,
  title={A theoretical and experimental investigation of temperature variation behind attenuating shock waves},
  author={Holbeche, TA and Spence, DA},
  journal={Proc. R. Soc. Lond. A},
  volume={279},
  pages={111--128},
  year={1964},
  publisher={The Royal Society London}
}

@article{susa2022flame,
  title={Flame image velocimetry: seedless characterization of post-reflected-shock velocities in a shock-tube},
  author={Susa, Adam J and Hanson, Ronald K},
  journal={Exp. Fluids},
  volume={63},
  pages={37},
  year={2022},
  publisher={Springer}
}

\end{document}